\documentclass[aps,pre,reprint,longbibliography,floatfix,superscriptaddress]{revtex4-2}
\usepackage{graphicx}
\usepackage{amsmath,amssymb,amsfonts}
\usepackage{hyperref}

\begin{document}

\title{Relaxation of saturated random sequential adsorption packings of discorectangles aligned on a line}
\author{Nikolai I. Lebovka}
\email[Corresponding author:]{lebovka@gmail.com}
\affiliation{Laboratory of Physical Chemistry of Disperse Minerals, F. D. Ovcharenko Institute of Biocolloidal Chemistry, NAS of Ukraine, Kyiv 03142, Ukraine}

\author{Mykhailo O. Tatochenko}
\email{tatochenkomihail@gmail.com}

\author{Nikolai V. Vygornitskii}
\email{vygornv@gmail.com}
\affiliation{Laboratory of Physical Chemistry of Disperse Minerals, F. D. Ovcharenko Institute of Biocolloidal Chemistry, NAS of Ukraine, Kyiv 03142, Ukraine}

\author{Yuri Yu. Tarasevich}
\email[Corresponding author: ]{tarasevich@asu.edu.ru}
\affiliation{Laboratory of Mathematical Modeling, Astrakhan State University, Astrakhan 414056, Russia}

\date{\today}

\begin{abstract}
Relaxation of the packing of elongated particles (discorectangles) aligned on a line was studied numerically. The aspect ratio (length-to-width ratio) for the discorectangles was varied within the range $\varepsilon \in [1;50]$. The initial jamming (saturated) state was produced using the basic variant of the random sequential adsorption (RSA) model with random positions and orientations of particles. The relaxation was performed by allowing rotational and translational diffusion motions of the particles wile their centers remained located on the line.  The effects of the aspect ratio $\varepsilon$ on the kinetics of relaxation, the orientation order parameter and the distribution function of the distances between nearest-neighbor discorectangles were analyzed. The transport properties of the resulting 1D systems were also analyzed by using the diffusion of a tracer particle (random walker) between the nearest-neighbor discorectangles. In the relaxed states the anomalous diffusion was observed having a hopping exponent $d_w>2$ dependent upon~$\varepsilon$.

\end{abstract}

\maketitle

\section{Introduction\label{sec:intro}}

The problem of one-dimensional (1D) random sequential adsorption (RSA) onto a line (the so-called car parking problem) has been studied in many works in the past~\cite{Renyi1963,Gonzalez1974,Evans1993,Talbot2000,Torquato2010,Krapivsky2010}. In this model, particles of given shape are placed randomly and sequentially without overlapping any previously placed particles, as their centers are located onto a line. Finally, after a sufficiently long period of deposition ($t \to \infty$), a jamming state is formed such that no additional particles can be added due to the absence of appropriate holes. In the basic variant of RSA, the absence of any relaxation (particle rotations or translations) is assumed.

The different 1D-RSA problems for particles with arbitrary shapes, e.g., segments (sticks), disks, ellipses, rectangles, discorectangles, etc.,  have been analyzed~\cite{Chaikin2006,Baule2017,Ciesla2020}. For elongated particles, this problem is commonly referred to as the ``Paris car parking problem''~\cite{Lebovka2020Paris}. For the parking of 1D segments of identical length, the kinetics of the RSA have been described analytically~\cite{Renyi1963,Gonzalez1974}. In the jamming state, the following value for the parking coverage (coverage of the line by segments) $\varphi_\text{R}=0.7475979203\dots$ (R\'{e}nyi's parking constant) has been obtained~\cite{Renyi1963,Clay2016}.

Similar problem have been analyzed for modified parking problems where there is the possibility of moving segments~\cite{Solomon1967}; where there are segments of different lengths~\cite{Ananevskii1999,Hassan2002,Hassan2002a}; packing of spins~\cite{Itoh1999}; the presence of arbitrary interaction potentials with finite ranges~\cite{Baule2019}; where there are  ``marking street'' effects~\cite{Seba2009}; or psychophysiological packing interactions (involving visual perception of space)~\cite{Seba2009}.

For RSA of particles of a given shape, the jamming coverage depends upon the aspect ratio $\varepsilon$ (the length-to-diameter ratio). In particular, for discorectangles it increased from $\varphi= 0.7476$ for $\varepsilon=1$ (disks), goes through a maximum $\varphi=0.781249 \pm 0.000020$) at $\varepsilon\approx 1.5$, and then decreases at  higher aspect ratios~\cite{Ciesla2020,Lebovka2020Paris}. This non-monotonic $\varphi (\varepsilon)$ dependence has been explained by the interplay between orientational degrees of freedom and excluded volume effects~\cite{Chaikin2006}.

The basic variant of the 1D-RSA problem produces out-of-equilibrium systems. By contrast, a diffusionally equilibrated system corresponds to a 1D Tonks gas~\cite{Thompson1988}. The gap distribution functions of totally irreversible and fully equilibrated 1D depositions of line segments demonstrate the presence of strong differences between these two systems~\cite{DOrsogna2004}. In a modified adsorption-desorption RSA model, the particles can be adsorbed with a rate of $k_+$, or desorbed with a rate of $k_-$. The properties of such packings depend on the ratio $k_+/k_-$~\cite{Krapivsky1994,Viot2005,Krapivsky2010}. In such a model, the systems reach equilibrium. However, their kinetics exhibit a succession of regimes before reaching equilibrium. To the best of our knowledge, the relaxation of 1D-RSA packings of elongated particles has never been studied before.

The present paper analyzes, numerically, the relaxation of RSA packings of elongated particles (discorectangles) on a line. The initial state was produced using the basic variant of the 1D-RSA problem~\cite{Lebovka2020Paris}. The relaxation was performed while accounting for the rotational and translational diffusion motions of the particles. The rest of the paper is constructed as follows. In Sec.~\ref{sec:methods}, the technical details of the simulations are described, all necessary quantities are defined, and some examples of the patterns are presented. Section~\ref{sec:results} presents our principal findings. The transport properties of 1D systems are also analyzed using the diffusion of a tracer particle (random walker) between the elongated particles along the line. Section~\ref{sec:conclusion} summarizes the main results.

\section{Computational model\label{sec:methods}}
\begin{figure}[!htbp]
\centering
\includegraphics[width=\columnwidth]{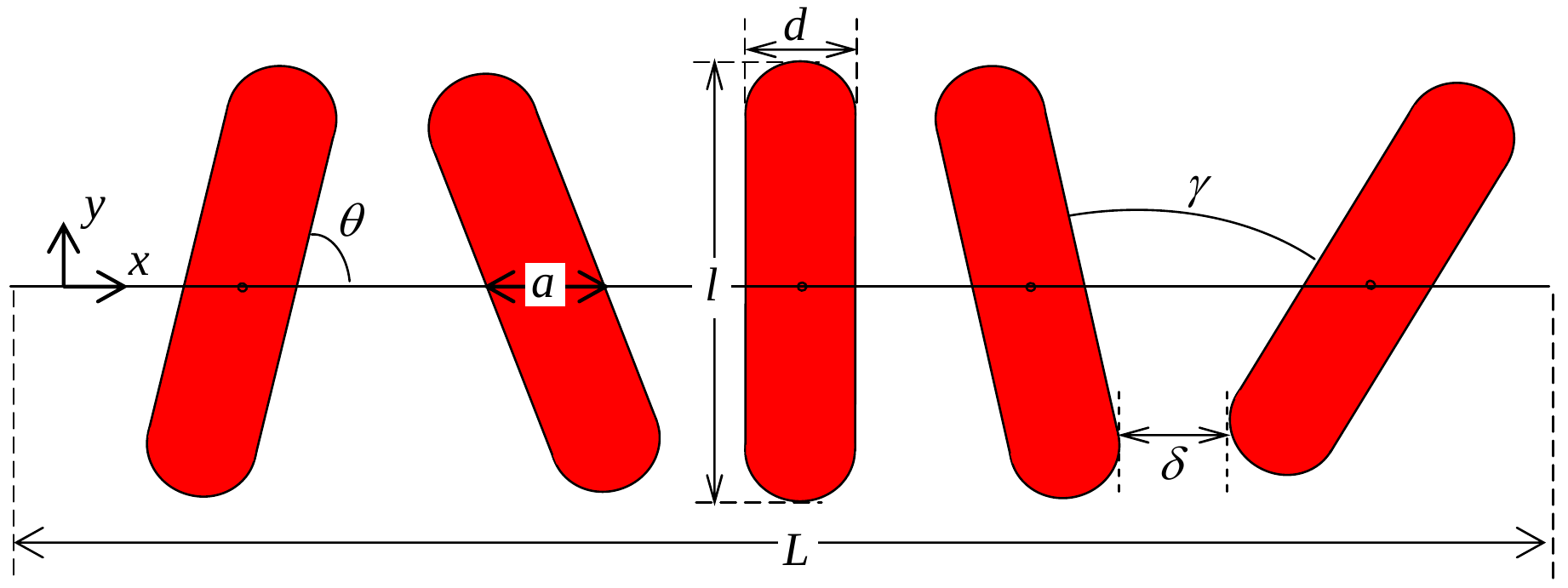}
\caption{Schematic picture of the RSA packing of elongated particles onto a line. The particles are hard discorectangles of length $l$ and width $d$. Intersections of the particles are forbidden. Each deposited particle covers a distance $a(\theta)$ on the line. Periodic boundary conditions are applied along the line ($x$-axis). Here, $L$ is the total length of the line, $\varepsilon=l/d$ is the aspect ratio, $\theta$ is the angle between the particle's long axis and the line, $\delta$ and $\gamma$ are, respectively, the shortest distance and the angle between nearest neighbor particles.
\label{fig:f01}}
\end{figure}

The initial state before relaxation was produced using an RSA model~\cite{Evans1993}. Hard discorectangles (rectangles with a semicircle at each of a pair of opposite sides) with length  $l$ and width $d$ were randomly and sequentially deposited onto a line. Overlapping of a particle with any previously deposited ones was forbidden, the orientations of particles were random and their centers were localized on the line (along the $x$-axis) (Fig.~\ref{fig:f01}).

A jamming state is where no additional particle can be added to the system due to the absence of any pores of appropriate size. To generate the jamming state a computationally efficient technique based on the tracking of local regions was employed (more detailed information can be found elsewhere~\cite{Lebovka2020Paris}). Problems for particles with aspect ratios $\varepsilon \in [1;50]$ were analyzed. To simplify presentation, all distances were measured in units of particle width. Periodic boundary conditions were used to minimize any finite-size effects, and typically, the length of the line was taken as $L = 32768\varepsilon$~\cite{Lebovka2020Paris}.
\begin{figure*}[!htbp]
	\centering	
\includegraphics[width=0.75\textwidth]{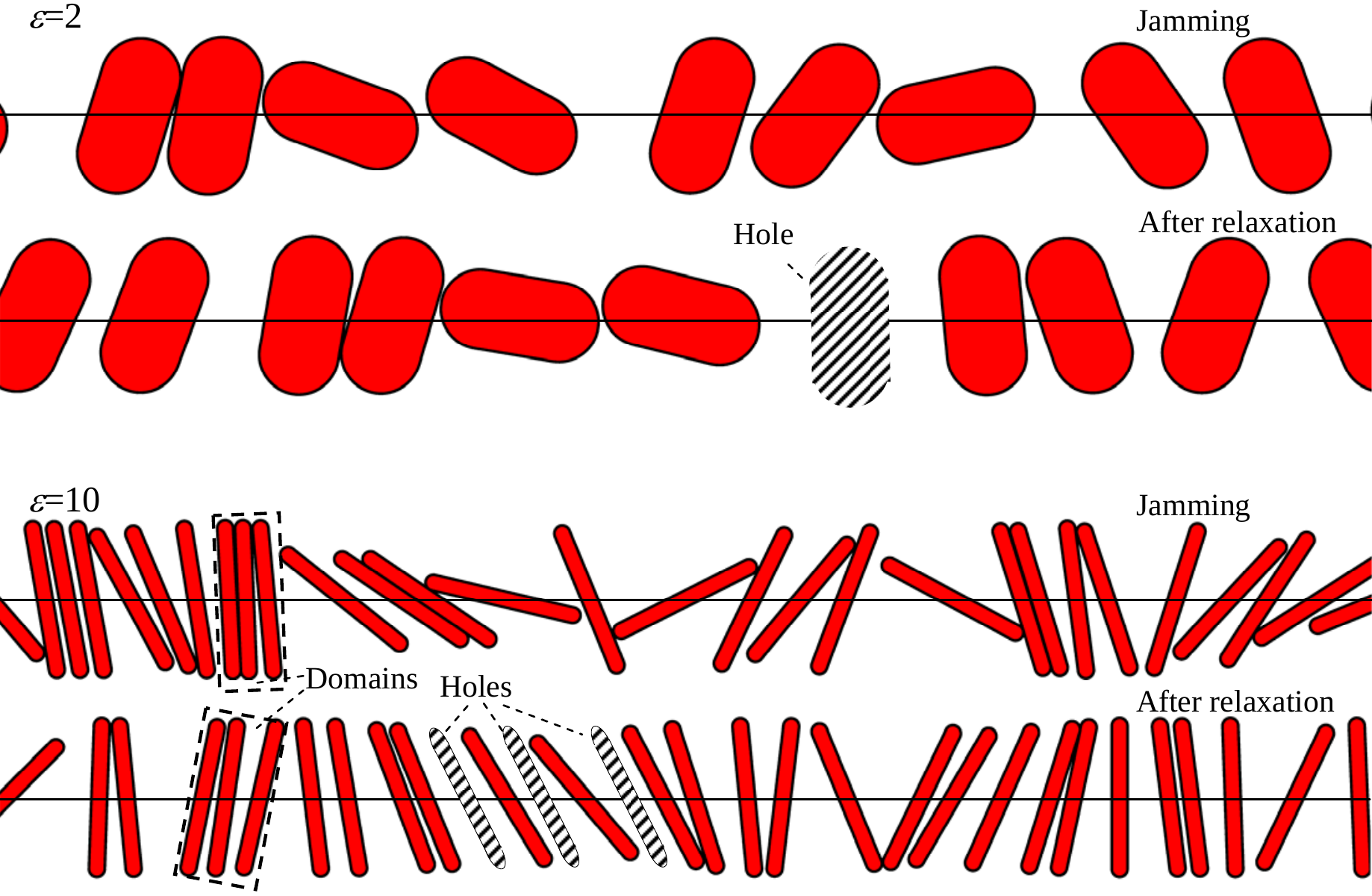}\\
	\caption{Examples of the patterns in a jamming state and after complete relaxation for particles with aspect ratios $\varepsilon=2$ and $\varepsilon=10$. The relaxation resulted in the appearance of rather large holes (shown as hatched areas). For relatively long particles ($\varepsilon=10$) domains of nearly parallel particles are also formed.
 \label{fig:f02} }
\end{figure*}

The number density was calculated as $\rho=N/L$, where $N$ is the total number of deposited particles. Each deposited particle covers a distance $a$ on the line, thus, the packing coverage was evaluated as $\varphi = \sum_i^N a_i/L$. The orientation of the particles was characterized by the order parameter defined as
\begin{equation}\label{eq:S}
  S=\overline{\cos 2\theta},
\end{equation}
where $\overline{\cdot\cdot\cdot}$ denotes the average, $\theta$ is the angle between the long axis of the particle and the line ($x$-axis) (Fig.~\ref{fig:f01}). Note that $S=1$ and $S=-1$ correspond to ideally oriented particles~--- along the $x$-axis or perpendicular to it, respectively.

The relaxation procedure was performed using the Monte Carlo (MC) approach as follows. At each step, an arbitrary particle was randomly chosen, and its rotational and translational diffusion motions were taken into account. The amplitudes of the Brownian motions were proportional to the square root of the corresponding diffusion coefficients. The values of these amplitudes were chosen to be small enough in order to obtain satisfactory acceptance of the MC displacements~\cite{Landau2014}. One MC time step ($\Delta t_\text{MC}=1$) corresponded to an  attempted single rotation and two translational displacements for each of all the particles in the system. The rotational diffusion coefficient was calculated as~\cite{Loewen1994}
\begin{equation}\label{eq:Dr}
  D_\text{r} = \frac{D_0(\ln \varepsilon+\gamma_\text{r})}{2\pi},
\end{equation}
where $D_0=k_\text{B} T/(\eta l)$, $k_\text{B} T$ is the thermal energy, and $\eta$ is the viscosity of the surrounding medium.

The centers of the particles were always fixed on the line ($x$-axis) and the translational diffusion coefficients were calculated as~\cite{Loewen1994}
\begin{subequations}\label{eq:Dpp}
\begin{align}
D_\parallel &= \frac{D_0 (\ln \varepsilon+\gamma_\parallel)}{2\pi},\\
D_\perp &= \frac{D_0 (\ln \varepsilon+\gamma_\perp)}{4\pi},
\end{align}
\end{subequations}
for the motions along and perpendicular to the direction of the long axis, respectively.

In the above Eqs.~\eqref{eq:Dr}--\eqref{eq:Dpp} the hydrodynamic end-correction factors $\gamma_\text{r}$, $\gamma_\parallel$, and $\gamma_\perp$ can be evaluated as~\cite{Loewen1994}
\begin{subequations}\label{eq:EndCorr}
\begin{align}
\gamma_\text{r} &=         -0.622 +0.917/\varepsilon-0.133/\varepsilon^2,\\
\gamma_\parallel &= -0.207 +0.980/\varepsilon-0.050/\varepsilon^2	,\\
\gamma_\perp &=      0.839 +0.185/\varepsilon+0.223/\varepsilon^2	.
\end{align}
\end{subequations}

The Brownian dynamics time increment was evaluated as~\cite{Patti2012}:
\begin{equation}\label{eq:DeltaT}
\Delta t_{B}=\frac{\mathcal{A}_i}{3} \Delta t_\text{MC},
\end{equation}
where $ \mathcal{A}_i $ is  the  acceptance coefficient for the $i$-th MC step~\cite{Landau2014}, and the total Brownian dynamics time was calculated  as
\begin{equation}
 t_{B}=\frac{ \Delta t_\text{MC} } {3}  \sum_{i=1}^{t_\text{MC}}\mathcal{A}_i,
\end{equation}
where $ t_\text{MC} $ is  the  MC time.

Time counting was started from the value of $t_\text{MC}=1$, being the initial moment (before relaxation), and the total duration required for the complete relaxation of the system in the equilibrium state was typically $10^4$--$10^5$ MC time units (more detailed information can be found elsewhere~\cite{Lebovka2019Sed,Lebovka2019Rel}). The completeness of relaxation was controlled by checking the changes in the order parameter, $S$ (Eq.~\eqref{eq:S}).

Figure~\ref{fig:f02} compares examples of the patterns in the jamming state and after complete relaxation for particles with aspect ratios $\varepsilon=2$ and $\varepsilon=10$. Relaxation resulted in the appearance of rather large holes (they are shown as hatched areas) suitable for further packing with new particles. For relatively long particles ($\varepsilon=10$), domains of nearly parallel particles are also formed both in the jamming state and after complete relaxation.  Therefore, the relaxation of RSA packings may result in considerable changes in the structure of such systems.

For each given value of $\varepsilon$, the computer experiments were averaged over $10-100$ independent runs. The error bars in the figures correspond to the standard errors of the means. When not shown explicitly, they are of the order of the marker size.

\section{Results and Discussion\label{sec:results}}

\begin{figure}[!thb]
\centering
\includegraphics[width=0.95\columnwidth]{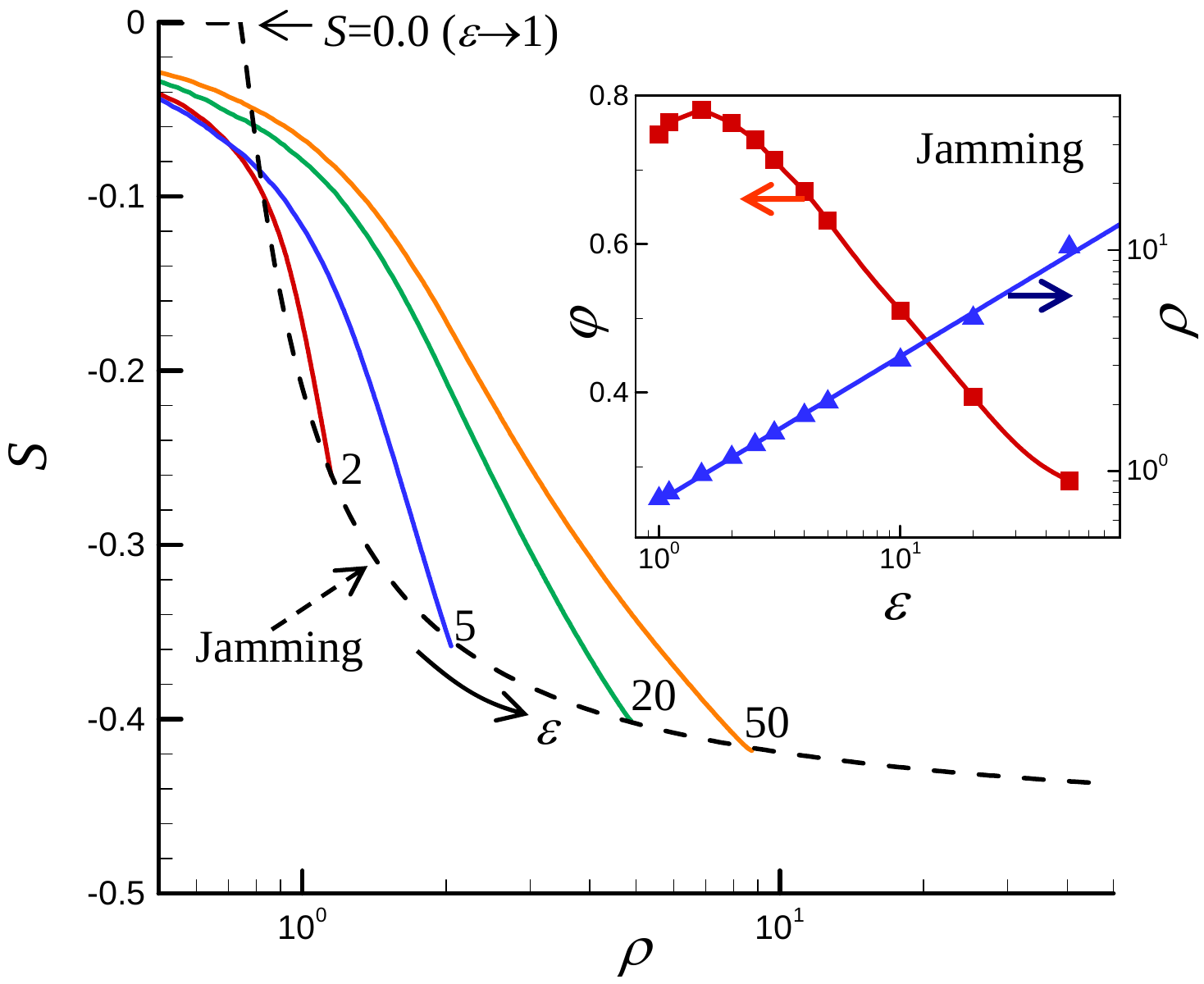}\\
\caption{Order parameter, $S$, versus the number density $\rho$ during the RSA process with different values of aspect ratio, $\varepsilon$. Dashed line corresponds to the $S(\rho)$ dependence in the jamming state. Inset shows the coverage $\varphi$ and density $\rho$ of packings in the jamming state versus $\varepsilon$.
 \label{fig:f03}}
\end{figure}

Figure~\ref{fig:f03} shows the order parameter, $S$, versus the number density, $\rho$, during the RSA process with different values of the aspect ratio, $\varepsilon$. At small values of $\rho$ the orientations of the particles were random ($S\approx 0$), while with increase of $\rho$, the order parameters decreased and reached their minimum values at the jamming state. This reflected the tendency toward particle ordering perpendicularly to the line, i.e., along the $y$-axis. The dashed line shows the $S(\rho)$ dependence in the jamming state. The inset to Fig.~\ref{fig:f03} compares the coverage $\varphi(\varepsilon)$ and the number density $\rho(\varepsilon)$ dependencies in the jamming state. The well-defined maximum in the $\varphi(\varepsilon)$ dependency  ($\varphi=0.7822 \pm 0.004$ at $\varepsilon\approx1.46$) can be explained by the competition between the orientational degrees of freedom and the excluded volume effects~\cite{Chaikin2006,Ciesla2020,Lebovka2020Paris}. The value $\rho$ grows continuously with $\varepsilon$ ($\rho \approx 0.73 \varepsilon^{0.66}$ ).

\begin{figure}[!htbp]
	\centering	
\includegraphics[width=0.9\columnwidth]{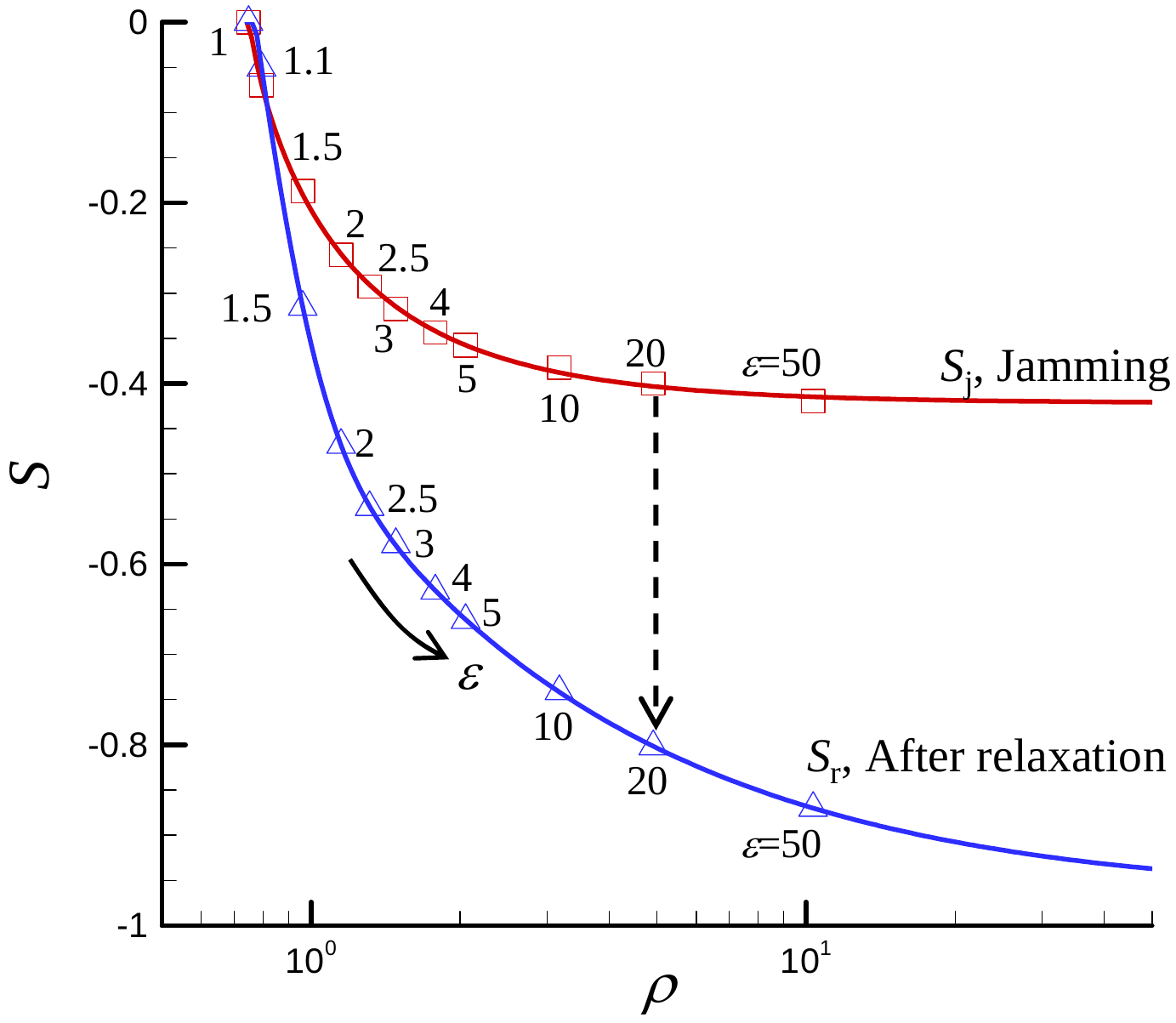}\\
\caption{Order parameter, $S$, versus number density, $\rho$, in the jamming state and after complete relaxation with different values of aspect ratio, $\varepsilon$. 	In the limit of infinitely thin sticks ($\varepsilon\to\infty$ and $\rho\to\infty$) the order parameter approached the values of $S\approx -0.450\pm 0.005$ and $S\approx -1$ in the jamming state and after complete relaxation, respectively. \label{fig:f04}}
\end{figure}

Figure~\ref{fig:f04} presents the order parameter, $S$, versus the number density, $\rho$, in both the jamming state ($S_\text{j}$) and after complete relaxation ($S_\text{r}$) with different values of the aspect ratio, $\varepsilon$. It is remarkable that, after complete relaxation, the systems had become orientationally ordered. The effects were more pronounced for particles  with large values of $\varepsilon$. Particularly, in the limit of infinitely thin sticks ($\varepsilon\to\infty$ and $\rho\to\infty$), the order parameter approached the values of $S\approx -0.450\pm0.005$ and $S\approx -1$ in the jamming state and after complete relaxation, respectively.

The transition to the more oriented state during relaxation was characterized by changes in the reduced order parameter defined as $S^*=(S-S_\text{j})/(S_\text{r}-S_\text{j})$. The value of $S^*$ varied between $0$ at $S=S_\text{j}$ (jamming state) and $1$ at $S=S_\text{r}$ (complete relaxation). The characteristic Brownian time $t_\text{B}^*$ was determined from the maximum in the first derivative curve $dS^*/dt_\text{B}$ (dashed line).
\begin{figure}[!htbp]
	\centering	
\includegraphics[width=0.9\columnwidth]{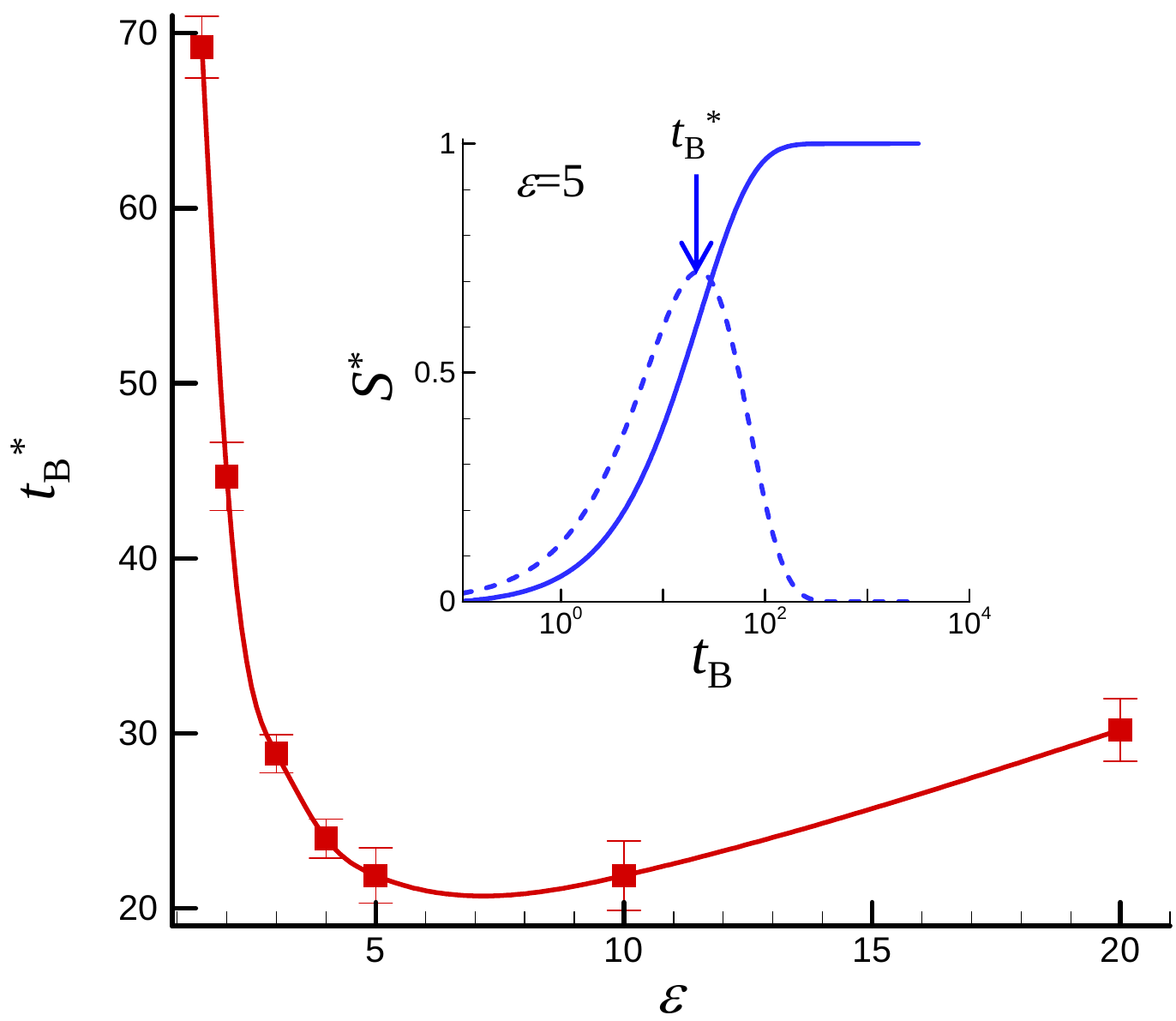}\\
	\caption{Characteristic Brownian time $t_\text{B}^*$ versus the aspect ratio, $\varepsilon$.
Inset shows example of reduced order parameter, $S^*=(S-S_\text{j})/(S_\text{r}-S_\text{j})$, versus the Brownian time, $t_\text{B}$, for aspect ratio $\varepsilon=5$. The value of $t_\text{B}^*$ was determined from the maximum of the derivative $dS^*/dt_\text{B}$ (dashed line). Here, $S_\text{j}$ and $S_\text{r}$ are the order parameters in the jamming state and after complete relaxation, respectively.
 \label{fig:f05} }
\end{figure}

Figure~\ref{fig:f05} presents the characteristic Brownian time $t_\text{B}^*$ versus the aspect ratio, $\varepsilon$. The inset shows an example of the reduced order parameter, $S^*$, versus the Brownian time $t_\text{B}$ for the aspect ratio $\varepsilon=5$. The slowest relaxation into the more oriented state was observed for particles with small aspect ratios. For this case, the value of $t_\text{B}^*$ went through a minimum at $\varepsilon\approx 7$. Such behavior may reflect competition between the orientational degrees of freedom and the excluded volume effects similar to that observed in the behavior of $\varphi(\varepsilon)$ (Fig.~\ref{fig:f04}). For elongated particles with large values of $\varepsilon$, slow relaxation may also be related to the formation of domains of nearly parallel particles (Fig.~\ref{fig:f02}).
\begin{figure}[!htbp]
	\centering	
\includegraphics[width=0.9\columnwidth]{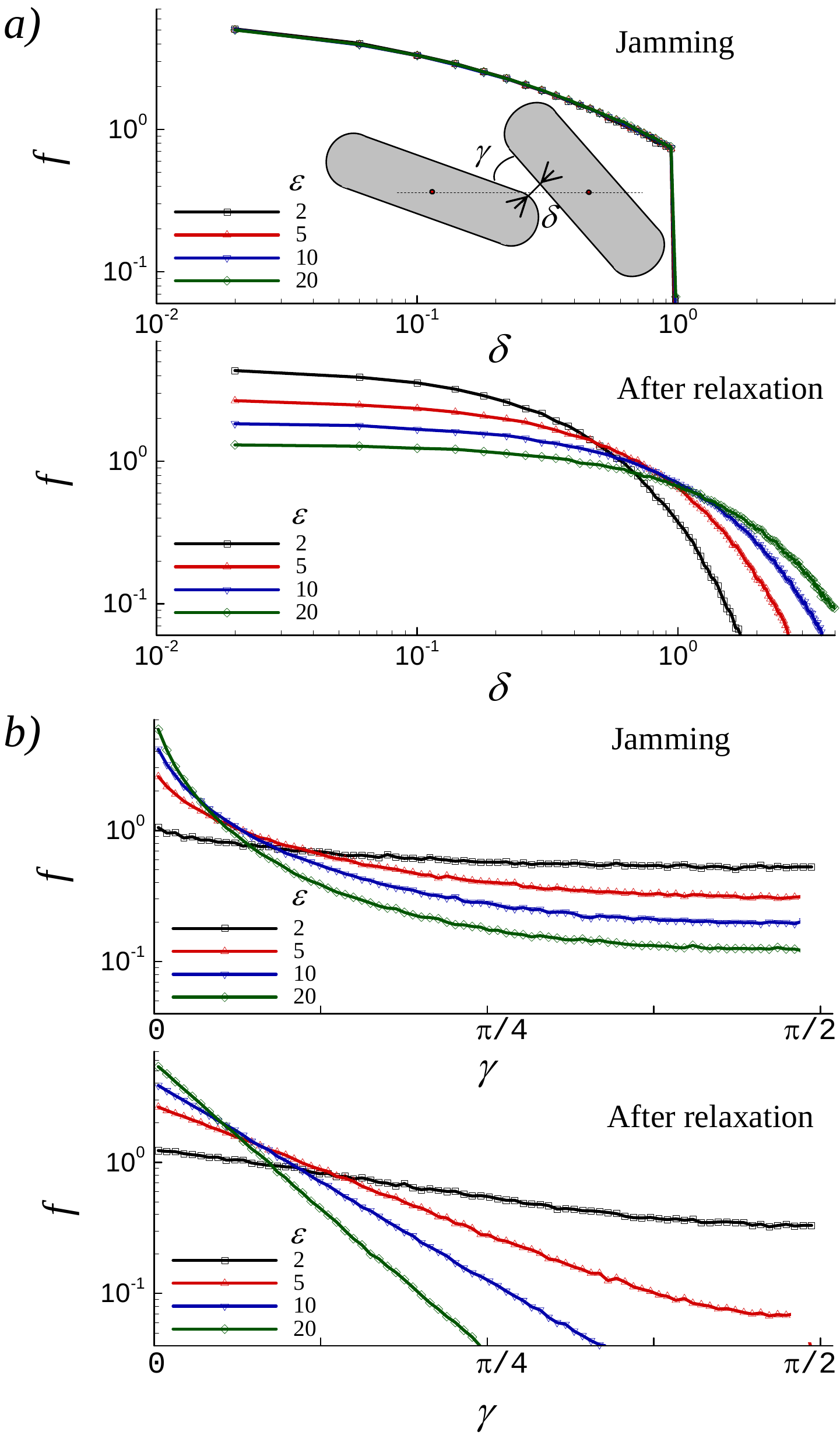}\\
	\caption{Distribution functions of the minimum distances $f(\delta)$ (a) and angles $f(\gamma)$ (b) between nearest-neighbor discorectangles with different aspect ratios, $\varepsilon$, in the jamming state and after complete relaxation. \label{fig:f06}}
\end{figure}

A deeper insight into differences in the systems in the jamming state and after complete relaxation can be obtained by analyzing the distribution functions of the minimum distances $f(\delta)$ (Fig.~\ref{fig:f06}a) and angles $f(\gamma)$ (Fig.~\ref{fig:f06}b) between nearest-neighbor discorectangles. In the jamming state, the distribution functions of the minimum distances $f(\delta)$ were non-zeros in the interval $0\leq\delta <1$ and were almost identical for different values of $\varepsilon$. Note that for disks  ($\varepsilon=1$) the obtained function $f(\delta)$ was similar to that observed earlier for the distribution function of the gaps between line segments~\cite{DOrsogna2004,Rawal2005}.

However, after complete relaxation, the function of $f(\delta)$ changed considerably (Fig.~\ref{fig:f06}a). The values of $\delta$ above 1 correspond to the formation of large holes suitable for the deposition of new particles. The effects were more significant for longer particles, with larger values of $\varepsilon$. In particular, the fractions of  $f(\delta)$ with $\delta$ above $1$ constituted $\approx 7.8\%$ and $\approx 50.3\%$ for $\varepsilon=2$ and for $\varepsilon=20$, respectively. A significant modification of the distribution functions has also been observed for thermally relaxed line segments (the Tonks gas)~\cite{DOrsogna2004}.
\begin{figure}[!htbp]
	\centering	
\includegraphics[width=0.9\columnwidth]{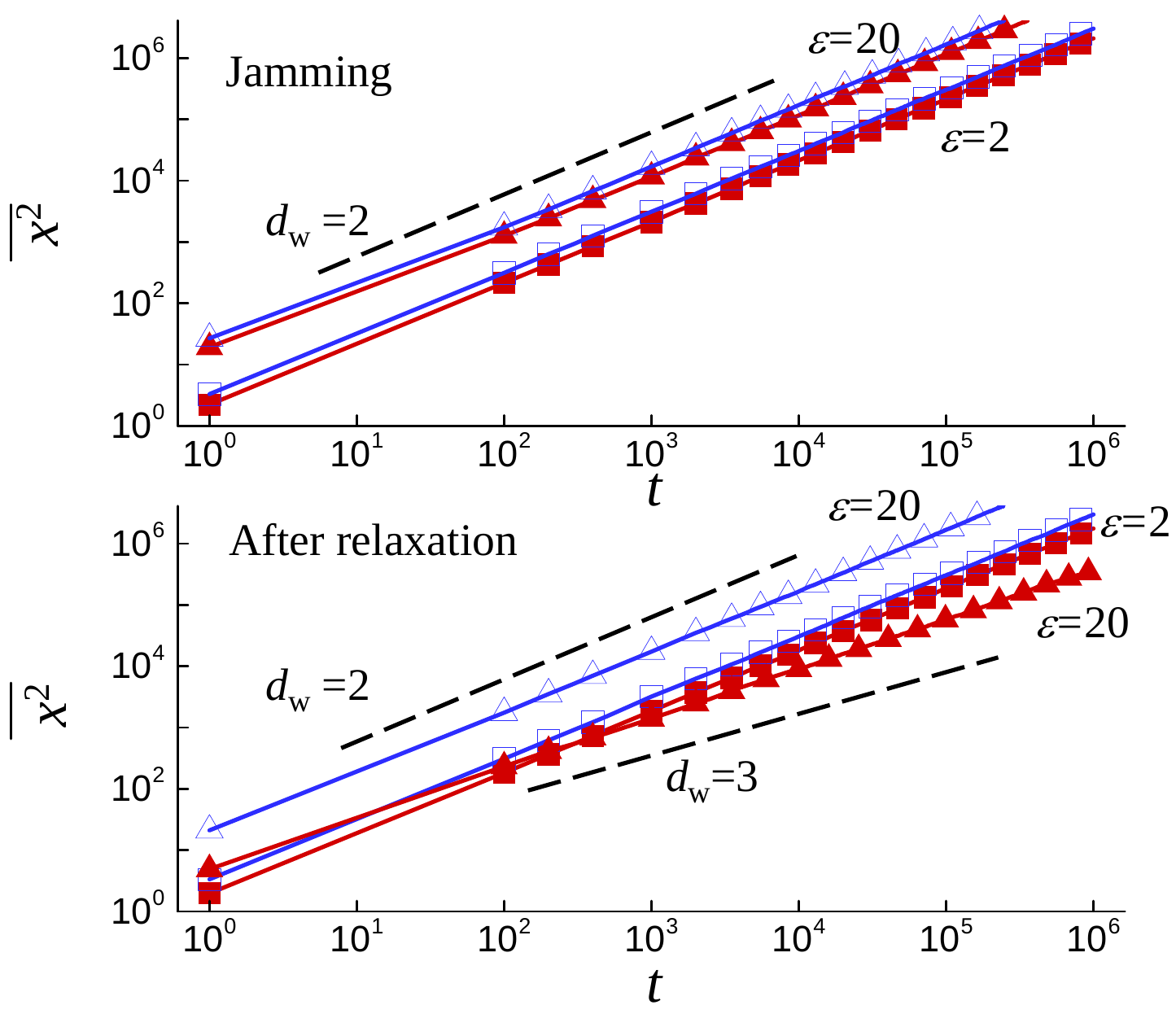}\\
	\caption{Mean square displacement of a tracer particle (random walker) along the line, $\overline{x^2}$, versus the time, $t$, (number of hopping steps) for systems in the jamming state and after complete relaxation for the aspect ratios $\varepsilon=2$ (squares) and $\varepsilon=20$ (triangles). The hopping particle jumps between the nearest-neighbor discorectangles with probability $p=\exp(-\delta/\lambda)$, where $\delta$ is the minimum distance between these discorectangles, and $\lambda$ is a diffusion distance parameter. The data are presented for $\lambda=1$ (closed symbols) and $\lambda=\infty$ (open symbols). Dashed lines represent the slopes for the hopping exponent $d_w=2$ and $d_w=3$.		
 \label{fig:f07} }
\end{figure}

The distribution functions of the angles $f(\gamma)$ between nearest-neighbor discorectangles were descending functions of $\gamma$ with maximum located at $\gamma=0$ (Fig.~\ref{fig:f06}b). Moreover, the values of the function $f(\gamma)$ were dependent on $\varepsilon$, being sharper for longer particles, and the effects were more pronounced for the systems that had undergone complete relaxation (Fig.~\ref{fig:f06}a). The observed effects reflected the formation of domains of nearly parallel particles (Fig.~\ref{fig:f02}). This transition to more oriented state during relaxation was accomplished with significant sharpening of the function~$f(\gamma)$.

Finally, the transport properties of quasi-one-dimensional system were analyzed using the diffusion of a tracer particle (random walker) between elongated particles along the line~\cite{Haus1987}. In the model, the tracer particle performed  random jumps (either left or right) to the nearest-neighbor discorectangle with a probability
\begin{equation}\label{eq:Prob}
p=\exp (-\delta/\lambda),
\end{equation}
where $\lambda$ is a diffusion distance parameter. Note that at $\lambda=\infty$ the jump probability is $p=1$ and the model is equivalent to unconstrained 1D diffusion.

In simulations, the mean square displacement of a tracer particle, $\overline{x^2}$, as a function of a discrete time, $t$, was analyzed. In the general case, in disordered systems diffusion law can be represented by the following equation~\cite{Metzler2000,Havlin2002,Masuda2017,Oliveira2019,Santos2019}
\begin{equation}\label{eq:Anomalous}
\overline{x^2}=(t/\tau)^{2/d_w},
\end{equation}
where the parameter $d_w$ (the hopping exponent) characterizes the deviation from normal diffusion, and $\tau$ is the characteristic diffusion time.
\begin{figure}[!htbp]
	\centering	
\includegraphics[width=0.9\columnwidth]{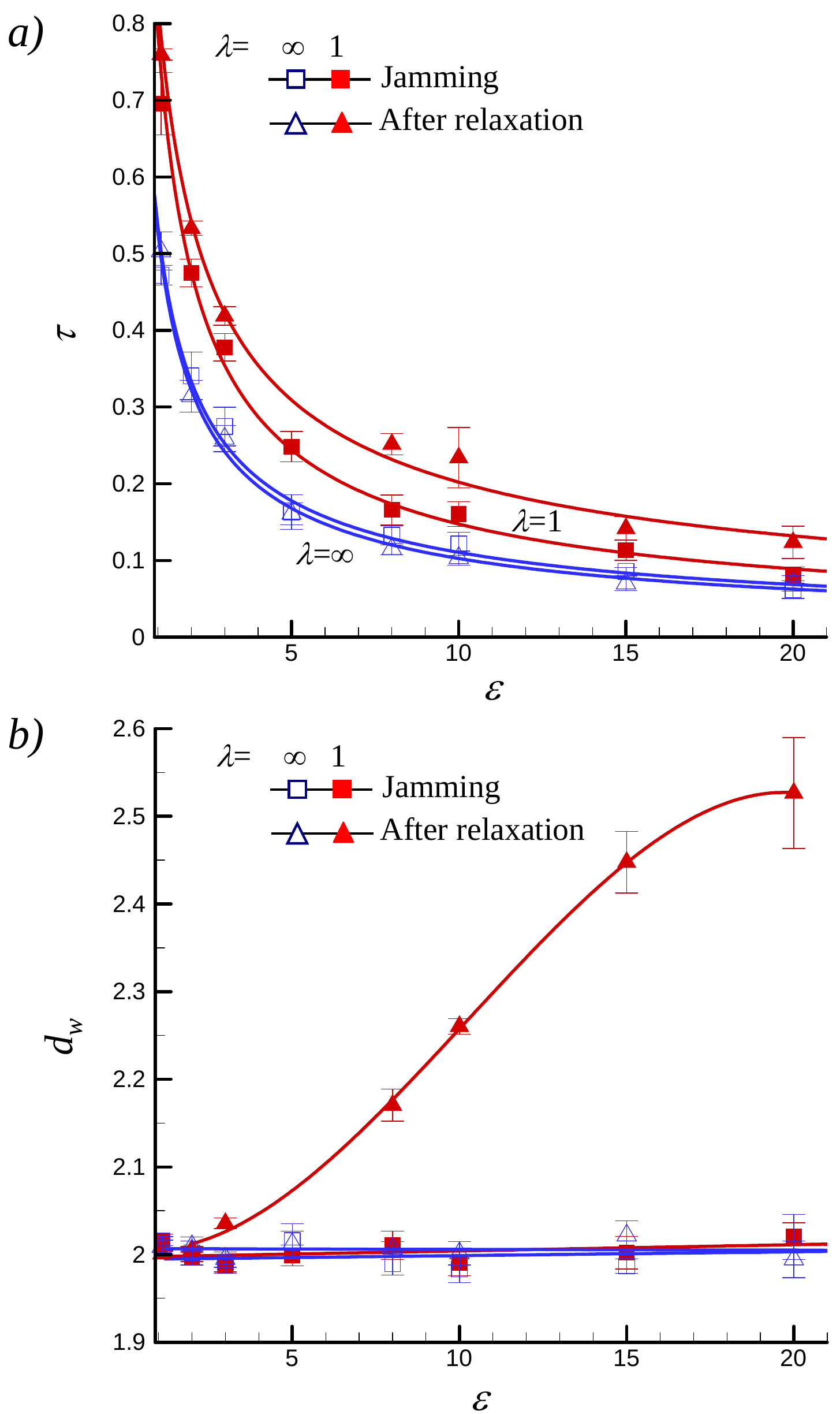}\\
	\caption{Characteristic diffusion time, $\tau$, (a) and hopping exponent $d_w$ (b) versus the aspect ratio $\varepsilon$ for systems in the jamming state (squares) and after complete relaxation (triangles) for $\lambda=1$ (closed symbols) and $\lambda=\infty$ (open symbols).
 \label{fig:f08} }
\end{figure}

For normal Fickian diffusion, $d_w=2$ and for 1D systems the diffusion coefficient is $D=(2\tau)^{-1}$. For the defined value of $D$ the d.c. electrical conductivity can also be established via the Einstein relation, $\sigma_{dc}=D e^2 n/(k_\text{B}T)$, where $e$ is the charge of an electron, and $n$ is the carrier density~\cite{Havlin2002}. The non-trivial hopping exponent $d_w > 2$ corresponds to the subdiffusion that occurs in disordered systems (e.g., gel-like or fractal)~\cite{Kubala2020}.

Figure~\ref{fig:f07} shows examples of the mean square displacement of a tracer particle (random walker) along the line, $\overline{x^2}$, versus the time, $t$, (number of hopping steps) for systems in the jamming state and after complete relaxation. The data are presented for the aspect ratios $\varepsilon=2$ (squares) and $\varepsilon=20$ (triangles), and for the diffusion distance parameters $\lambda=1$ (closed symbols) and $\lambda=\infty$ (open symbols). The dashed lines represent the slopes for the hopping exponent $d_w=2$ and $d_w=3$.	

No effects of anomalous diffusion were observed for the systems with $\lambda=\infty$, i.e. for  normal 1D diffusion. In the jamming state, normal diffusion with the hopping exponent $d_w=2$ was always observed for all values of  $\varepsilon$. However, after complete relaxation, the effects of anomalous diffusion with  $d_w>2$ were evident and could be observed for $\lambda=1$.

Figure~\ref{fig:f08} presents examples of the characteristic diffusion time, $\tau$, (a) and the hopping exponent $d_w$ (b) versus the aspect ratio $\varepsilon$ for  systems in the jamming state (squares) and after complete relaxation (triangles) for $\lambda=1$ (closed symbols) and $\lambda=\infty$ (open symbols). The characteristic diffusion time, $\tau$, decreased with increase of $\varepsilon$ (Fig.~\ref{fig:f08}). For  normal diffusion, it corresponded to the increase of the diffusion coefficient $D=1/(2\tau)$. This increase of $D$ with $\varepsilon$  may reflect the increase in the length of discorectangles.

A significant increase in value of the hopping exponent $d_w$ was observed after relaxation. This behavior is in correspondence with the observed effects of  relaxation on the distribution functions $f(\delta)$ (Fig.~\ref{fig:f06}a).

\section{Conclusion\label{sec:conclusion}}

Numerical studies of relaxation packings of elongated particles (discorectangles) on a line were performed. The initial jamming state was produced using the basic variant of the 1D-RSA problem~\cite{Lebovka2020Paris}. During the RSA deposition, the orientations of the particles were selected at random. In the RSA state, a tendency towards particle ordering perpendicularly to the line ($x$-axis) was observed. This self-organization was more pronounced for particles with large values of aspect ratio $\varepsilon$. Our findings provide insight into how the value of $\varepsilon$ affects the transition from the jamming to the relaxed state. This transition into the more oriented state was accompanied by the formation of domains of nearly parallel particles, and the appearance of rather large holes, suitable for the placement of additional particles. The relaxed systems also demonstrated anomalous transport properties when tested using the diffusion of a tracer particle (random walker).

\acknowledgments
We acknowledge funding from the National research foundation of Ukraine, Grant No.~2020.02/0138 (M.O.T., N.V.V.), the National Academy of Sciences of Ukraine, Project Nos.~7/9/3-f-4-1230-2020,~0120U100226 and~0120U102372/20-N (N.I.L.), and funding from the Foundation for the Advancement of Theoretical Physics and Mathematics ``BASIS'', Grant No. 20-1-1-8-1 (Y.Y.T.).


\bibliography{Lebovka2021_1dRelax}  %

\end{document}